\documentclass[prl,twocolumn,superscriptaddress,amsmath,amssymb]{revtex4}
\usepackage{graphicx}
\usepackage{bm}
\usepackage{amsmath,amssymb}
\usepackage{color}
\include{Qcircuit}

\begin{document}
\title{On the hardness of classically simulating the one clean qubit model} 
\author{Tomoyuki Morimae}
\email{morimae@gunma-u.ac.jp}
\affiliation{ASRLD Unit, Gunma University, 1-5-1 Tenjin-cho Kiryu-shi
Gunma-ken, 376-0052, Japan}
\author{Keisuke Fujii}
\email{keisukejayorz@gmail.com}
\affiliation{The Hakubi Center for Advanced Research, Kyoto University,
Yoshida-Ushinomiya-cho, Sakyo-ku, Kyoto 606-8302, Japan}
\affiliation{Graduate School of Informatics, Kyoto University,
Yoshida Honmachi, Sakyo-ku, Kyoto 606-8501, Japan}
\author{Joseph F. Fitzsimons}
\email{joseph_fitzsimons@sutd.edu.sg}
\affiliation{Singapore University of Technology and Design,
20 Dover Drive, Singapore 138682}
\affiliation{Center for Quantum Technologies,
National University of Singapore, Block S15, 3 Science Drive 2, Singapore 117543}

\date{\today}
            
\begin{abstract}
Deterministic quantum computation with one quantum bit (DQC1) 
[E. Knill and R. Laflamme, Phys. Rev. Lett. {\bf81}, 5672 (1998)]
is a model of quantum computing
where the input is restricted to containing a single qubit in a pure state 
and has all other qubits in a completely-mixed
state. Only the single pure qubit is measured at the end of the computation.
While it is known that DQC1 can efficiently solve
several problems for which no known classical efficient algorithms exist, the question of whether DQC1 is really more powerful than classical computation remains open. In this paper, we introduce a slightly modified version of DQC1, 
which we call DQC1$_k$, where $k$ output qubits are measured,
and show that DQC1$_{k}$ cannot be classically efficiently simulated for any $k\geq3$ unless
the polynomial hierarchy collapses at the third level.
\end{abstract}

\pacs{03.67.-a}
\maketitle  

\begin{figure}[htbp]
\begin{center}
\[
\Qcircuit @C=0.5em @R=0.5em {
&\lstick{a)}&\lstick{\ket{0}}&\qw & \multigate{4}{~U~} & \qw &\meter & \push{~~~~~~~~} & \lstick{b)} &\lstick{\ket{0}}&\gate{H} & \ctrl{1} & \gate{H} &\meter\\
&\push{~~~~~}&\lstick{\frac{I}{2}}&\qw & \ghost{~U~} & \qw & \qw & ~~~~~ &\push{~~~~~}&\lstick{\frac{I}{2}}&\qw &  \multigate{3}{~U~} & \qw & \qw \\
&&\lstick{\frac{I}{2}}&\qw & \ghost{~U~} & \qw & \qw & ~~~~~ &&\lstick{\frac{I}{2}}&\qw & \ghost{~U~} & \qw & \qw \\
&&\push{\vdots} && & \push{~\vdots}& && ~~~~~ &\push{\vdots}& & ~~~ & \push{~\vdots}&  \\
&&\lstick{\frac{I}{2}}&\qw & \ghost{~U~} & \qw & \qw & ~~~~~ && \lstick{\frac{I}{2}}&\qw & \ghost{~U~} & \qw & \qw
}
\]
\end{center}
\caption{ The one clean qubit model. a) A circuit of the DQC1 model, b) a circuit of the DQC1 model whose output can evaluate Tr($U$),
for which classical efficient algorithm is not known. All measurements are assumed to be in the computational basis.
} 
\label{DQC1circuit}
\end{figure}
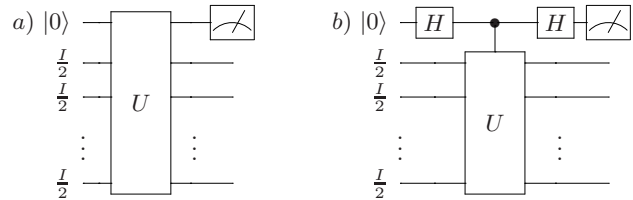

While large scale universal quantum computers may be many years off, several intermediate models of quantum computation have been discovered which may prove significantly easier to implement in practice. These intermediate models of computation do not offer the full potential of universal quantum computation, but are nonetheless believed to be hard to simulate classically. Motivated by nuclear magnetic resonance (NMR) 
quantum information processing,
Knill and Laflamme~\cite{KL} proposed a restricted model
of quantum computing, known as deterministic quantum computation with one quantum bit
(DQC1), or the one clean qubit model. As is shown in Fig.~\ref{DQC1circuit}a, a DQC1 circuit consists of the input state in a highly mixed state which is acted upon by a number of quantum gates polynomial in the size of the input, followed by a computational basis measurement of the first qubit. The initial state of the system is given by $\rho_{in}^{(n+1)}\equiv|0\rangle\langle0|\otimes \left(\frac{I}{2}\right)^{\otimes n}$ where $I$ is the two dimensional 
identity operator. We call this state ``the highly-mixed input state".
Naively, one might expect this model to be easy to simulate classically, since any time evolution of a single qubit state can be 
trivially simulated efficiently by a classical computer and the completely-mixed state seems to lack
``quantumness", due to a lack of entanglement and discord.
However, the surprising result of Ref.~\cite{KL}
is that DQC1 can efficiently solve certain problems for which no efficient classical algorithms are known. For example, the quantum circuit represented Fig.~\ref{DQC1circuit}b can be used to estimate the normalized trace of any $n$-qubit unitary operator $U$~\cite{KL}, a problem which is complete for the class of decision problems answerable within this model with bounded error~\cite{earlier}. This is possible due, in part, to the fact that while entanglement remains bounded in DQC1 circuits, the interaction between the mixed register and the pure qubit leads to the presence of significant non-classical correlations~\cite{Datta2,Datta3}.

While it does not seem that this model supports universal quantum computation~\cite{Ambainis}, it can efficiently solve problems for which no efficient classical algorithm is known, such as spectral density estimation~\cite{KL}, testing integrability~\cite{Poulin}, calculation of fidelity decay~\cite{Poulin2},
and approximation of the Jones and HOMFLY polynomials~\cite{SS,Passante,JW}.
An algorithm for approximating an invariant of 3-manifolds was also proposed~\cite{3mani}.
Since the estimation of the normalized trace of a unitary matrix seems to be hard for
classical computers~\cite{Datta}, and DQC1 does not seem to be universal, the DQC1 model appears to represent a model of computation which is intermediate classical and universal quantum computation.

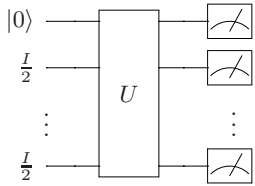
\begin{figure}[htbp]
\begin{center}
\[
\Qcircuit @C=1em @R=0.5em {
&\lstick{\ket{0}}&\qw & \multigate{3}{~U~} & \qw &\meter\\
&\lstick{\frac{I}{2}}&\qw & \ghost{~U~} & \qw & \meter \\
&\push{\vdots} && && \push{~\vdots}& \\
&\lstick{\frac{I}{2}}&\qw & \ghost{~U~} & \qw & \meter
}
\]
\end{center}
\caption{
A circuit of the DQC1$_{n+1}$.
} 
\label{DQC1variant_1and2}
\end{figure}

Although the presence of non-trivial quantum discord and entanglement within DQC1 circuits is seen as an indication that the trace
of $n$-qubit unitary matrix is classically hard to compute~\cite{Datta2,Datta}, ruling out many classical approaches to simulation~\cite{Datta3},
it is an open problem whether DQC1 really represents a more powerful model than purely classical computation.
In this paper, we show that a 
slightly modified version
of DQC1, which we call DQC1$_{n+1}$, is hard to be classically
efficiently simulated unless the polynomial hierarchy (PH) 
collapses at the third level. The DQC1$_{n+1}$ model is equivalent to the DQC1 model except
that all $n+1$ output qubits (instead of the single qubit) are measured at the end of the computation (see Fig.~\ref{DQC1variant_1and2}). We shall show that this result hardness holds even if we consider measurements on only a constant number of output qubits. Thus we demonstrate an inextricable link between the computational hardness of simulating circuits with one clean qubit and a widely accepted conjecture from computational complexity theory.
PH is a natural way of classifying the complexity 
of problems (languages) beyond the usual NP (nondeterministic polynomial time,
which includes "traveling salesman" and "satisfiability" problems).
It is strongly believed in computer science that NP includes
non-polynomial-time problems. Similarly, there is a weaker, but still
solid belief that PH does not collapse. 

Our argument is based on the seminal results by
Terhal and DiVincenzo~\cite{TD}, Bremner, Jozsa, and Shepherd~\cite{BJS},
and Aaronson and Arkhipov~\cite{AA}.
These papers introduced models of quantum computing that are not universal
but cannot be classically efficiently simulated unless
some plausible assumptions in computer science are violated. The essential idea behind all of these results is that
when post-selected some superficially naive circuits can simulate universal quantum computation (BQP) or postselected universal quantum computation ($\mbox{post-BQP}$). In this context, postselection means the (fictitious) ability to project onto a specific branch of the wavefunction with unit probability.
Therefore, if the probability distributions of the outputs of
such naive circuits can be classically efficiently simulated,
this means that the classical computer with a postselection
can also efficiently simulate
BQP or post-BQP circuits, which violate certain strongly believed conjectures in computer science.

Using such an approach, Terhal and DiVincenzo~\cite{TD} showed that it is hard to
simulate quantum circuits with depth four.
They derived the result by noticing the fact that  
non-adaptive Gottesman-Chuang quantum circuits~\cite{GC} can be written
with depth-four circuits.
Bremner, Jozsa, and Shepherd~\cite{BJS} showed that a class of
quantum circuits known as Instantaneous Quantum Polynomial-time
(IQP) circuits~\cite{FMIQP} cannot be classically efficiently simulated unless
PH collapses at the third level.
An $n$-qubit IQP circuit is a circuit that consists of the input state $|0\rangle^{\otimes n}$, a polynomial number of mutually commuting quantum gates, and computational-basis measurement on all $n$ output qubits
at the end of the computation. Each of the quantum gates in this model can be assumed to be of the form $D(\theta_j,S_j)\equiv \exp\big[i\theta_j\bigotimes_{k=1}^nX_k^{s_j^k}\big]$,
where $\theta_j\in {\mathbb R}$, $X_k$ is the Pauli $X$ 
operator acting on $k$th qubit,
and
$S_j\equiv(s_j^1,...,s_j^n)\in\{0,1\}^n$. What Bremner, Jozsa, and Shepherd showed was that postselected IQP circuits can simulate $\mbox{post-BQP}$ circuits. By combining this with a previous result of Aaronson~\cite{Aaronson}, that
$\mbox{post-BQP}=\mbox{PP}$, they concluded that if IQP circuits can be classically efficiently simulated, PH
collapses at the third level.
Finally, Aaronson and Arkhipov~\cite{AA} showed a hardness proof for classical simulations given of non-interacting bosons~\cite{Aaronson2}
which made use of probabilistic entangling gates due to Knill, Laflamme, and Milburn~\cite{KLM}, to show that such systems can simulate BQP circuits, and hence $\mbox{post-BQP}$ circuits, if the postselection is possible on the occupancy of modes\footnote{Aaronson and Arkhipov also gave a direct proof of the result without using the postselection argument~\cite{AA}}.

Here we make use of a similar approach. We show that
if postselections of the measurement results are possible, 
DQC1$_{n+1}$ can simulate $\mbox{post-BQP}$ circuits. 
Then, by using the fact that $\mbox{post-BQP}=\mbox{PP}$, we conclude that
if DQC1$_{n+1}$ can be efficiently classically simulated, the polynomial 
hierarchy collapses at the third level.

Before proceeding to the proof of our results, we first clarify what we mean by classically efficient simulation.
We adopt the definition used by Bremner, Jozsa, and Shepherd~\cite{BJS}.
(They call the definition ``weakly" simulatable with multiplicative error
$c\ge1$. But in this paper we sometimes omit the word ``weakly"
for simplicity, since we consider only this definition.)
For any uniform family of circuits $\{C_w\}$~\cite{uniform} let $P_w$ be the output probability
distribution of $C_w$.
Let us assume that we perform computational-basis measurements on the $k$ output qubits of each of an uniform family of quantum circuits.
Let $P_w(m_1,...,m_k)$ be the probability of obtaining measurement 
result $(m_1,...,m_k)\in\{0,1\}^k$.
We say that the family is (weakly) simulable with multiplicative error $c\ge1$ 
if for any marginal distribution $P_w(x_1,...,x_r)$ of $P_w(m_1,...,m_k)$
there exists a (family of) probability distributions $P_w'$ such that
it can be sampled classically in a polynomial time
and for all variables and $w$ we have
\begin{eqnarray*}
\frac{1}{c}P_w(x_1,...,x_r)\le P_w'(x_1,...,x_r)\le c P_w(x_1,...,x_r).
\end{eqnarray*}
(Note that the circuit size is included in the input complexity.)

We now give a more precise definition of our model. A DQC1$_k$ circuit consists of the highly-mixed input state $\rho_{in}^{(n+1)}$, to which a polynomial number of quantum gates chosen from a discrete approximately universal gate set are applied, followed by measurement of $k$ labeled qubits in the computational basis. Clearly, then, DQC1 is equivalent to the special case of $k=1$. At the opposite extreme, as is shown in Fig.~\ref{DQC1variant_1and2}, the DQC1$_{n+1}$ model is equivalent to the DQC1 model except that all output qubits are measured at the end of the computation.

With this definition in place, we can now make a precise statement of our main result: if DQC1$_{n+1}$ is classically simulable with multiplicative
error $1\le c< \sqrt{2}$, then PH collapses at the third level. Further, this result can be refined to show that the same result holds if DQC1$_{n+1}$ is replaced by DQC1$_{3}$.

In order to prove these claims, we begin by clarifying 
definition of postselected computation classes~\cite{BJS}.
A language $L$ is in the class 
post-$X$ if and only if there exists an error tolerance $0<\delta<\frac{1}{2}$
and a uniform family of postselected circuits (in the type specified 
by $X$, such as BQP, IQP, etc.) with a specified single (qu)bit output register $O_w$ (for the
$L$-membership decision problem) and a specified multi-(qu)bit ``postselection register" $P_w$\footnote{While post-BQP and post-BPP are usually defined with only a two-dimensional output register, their power is not increased by changing the size of this register} such that:
\begin{itemize}
\item[1.]
if $w\in L$ then $\mbox{Prob}(O_w=1|P_w=0...0)\ge \frac{1}{2}+\delta$ and
\item[2.]
if $w\notin L$ then $\mbox{Prob}(O_w=1|P_w=0...0)\le \frac{1}{2}-\delta$.
\end{itemize}
For post-BQP, the specification $X$ is the set of universal quantum circuits start with input fixed as $|0\rangle^{\otimes n}$. Similarly, for post-BPP, 
it is the set of randomised classical circuits with input fixed in the zero state, and for post-DQC1$_k$, it is the set of DQC1$_k$ circuits,
i.e., the highly-mixed input state $\rho_{in}^{(n+1)}$, 
a polynomial number of quantum gates on it, and
the measurement of $k$ labeled qubits in the computational basis. 

As discussed earlier, it has previously been established that $\mbox{post-BQP}=\mbox{PP}$~\cite{Aaronson}.
Furthermore, a non-trivial containment for post-BPP is known. Let PH denote the polynomial hierarchy: the union of
an infinite hierarchy of classes $\Sigma_k \text{P}$, $\Delta_k \text{P}$ and $\Pi_k \text{P}$ for $(k=0,1,2,...)$ where $\Sigma_0 \text{P}= \Delta_0 \text{P} = \Pi_0 \text{P} = \text{P}$ and 
$\Sigma_{k+1}=\mbox{NP}^{\Sigma_k \text{P}}$, $\Delta_{k+1}\text{P}=\mbox{P}^{\Sigma_k \text{P}}$ and $\Pi_{k+1}=\mbox{coNP}^{\Sigma_k \text{P}}$. It is known that ${\mbox{P}}^{\mbox{post-BPP}}\subseteq\Delta_3 \text{P}$~\cite{BJS},
and $\mbox{PH}\subseteq {\mbox{P}}^{\mbox{PP}}$~\cite{Toda}.

\begin{figure}[htbp]
\begin{center}
\[
\Qcircuit @C=1em @R=0.5em {
&\lstick{\ket{0}}&\qw & \multigate{4}{~W~} & \qw &\meter\\
&\lstick{\frac{I}{2}}&\qw & \ghost{~W~} & \gate{V_1} & \meter \\
&\lstick{\frac{I}{2}}&\qw & \ghost{~W~} & \gate{V_2} & \meter \\
&\push{\vdots} && && \push{~\vdots}& \\
&\lstick{\frac{I}{2}}&\qw & \ghost{~W~} & \gate{V_n} & \meter
}
\]
\end{center}
\caption{
A circuit of the DQC1$_{n+1}$ model which implements postselected measurement based quantum computation.
} 
\label{W}
\end{figure}
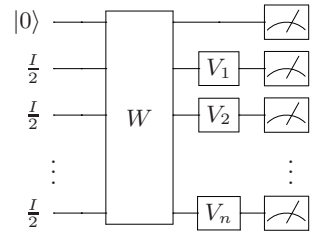

We now proceed to show that $\mbox{post-DQC1}_{n+1}=\mbox{post-BQP}$.
First, $\mbox{post-DQC1}_{k}\subseteq\mbox{post-BQP}$ is easy to show for any $k$ since the mixed-state input can be simulated 
with BQP circuits by adding ancilla qubits and entangling them 
qubits used in the computation, to leave the reduced system in the same mixed state as used for DQC1$_k$. Hence $\mbox{post-DQC1}_{n+1}\subseteq\mbox{post-BQP}$. Next we show the opposite containment, $\mbox{post-DQC1}_{n+1}\supseteq\mbox{post-BQP}$.
Let us consider an $n$-qubit cluster state
$|G\rangle$, which  is a universal resource state for the measurement-based quantum
computing~\cite{MBQC}.
Consider the DQC1$_{n+1}$ circuit 
(shown in Fig.~\ref{W})
with the gate 
$(I\otimes\bigotimes_{j=1}^nV_j)W$,
where $V_j$ is any single-qubit unitary gate, and
$W\equiv X\otimes |G\rangle\langle G|
+I\otimes (I^{\otimes n}-|G\rangle\langle G|)$. The unitary operator $W$
can be uniformly generated since the $m$-controlled Toffoli gate,
$|0\rangle\langle0|^{\otimes m}\otimes X+
(I^{\otimes m}-|0\rangle\langle0|^{\otimes m})\otimes I$,
can be uniformly generated without requiring any ancilla qubits~\cite{Barenco},
and a unitary transformation that takes $|0...0\rangle$ to a cluster state
can be uniformly generated without requiring any ancilla qubits.
The state after the application of $W$ is
$
\frac{1}{2^n}|1\rangle\langle1|\otimes|G\rangle\langle G|+
\frac{1}{2^n}|0\rangle\langle0|\otimes (I^{\otimes n}-|G\rangle\langle G|).
$

Therefore, if we postselect on the first qubit being in state $|1\rangle$,
we obtain the $n$-qubit cluster state $|G\rangle$. Generally measurement based computation requires that adaptive single qubit measurements be made based on previous measurement outcomes in order to achieve deterministic quantum computation. However, since we are allowing for postselection, by postselecting measurement outcomes on one for non-output qubits, it is possible to fix the measurement bases before beginning the computation, and hence postselection combined with the circuit in Fig 3 is sufficient to implement postselected BQP circuits with only polynomial overhead, and so $\mbox{post-DQC1}_{n+1}\supseteq\mbox{post-BQP}$.

Finally, we show that the classical simulability of DQC1$_{n+1}$ leads to
$\mbox{post-DQC1}_{n+1}\subseteq\mbox{post-BPP}$, following the argument of Bremner, Jozsa and Shepherd.
Assume that 
DQC1$_{n+1}$ is simulable with multiplicative error
$c\ge1$. 
Let $(O_w,P_w)$ and $(O_w',P_w')$ denote the output and postselection registers for
postselected DQC1$_{n+1}$ circuits and postselected
randomized classical circuits respectively.
Then, 
\begin{eqnarray*}
\mbox{Prob}(O_w'=x|P_w'=0...0)&=&\frac{\mbox{Prob}(O_w'=x,P_w'=0...0)}
{\mbox{Prob}(P_w'=0...0)}\\
&\ge& \frac{1}{c^2}\mbox{Prob}(O_w=x|P_w=0...0),
\end{eqnarray*}
and
\begin{eqnarray*}
\mbox{Prob}(O_w'=x|P_w'=0...0)&=&\frac{\mbox{Prob}(O_w'=x,P_w'=0...0)}
{\mbox{Prob}(P_w'=0...0)}\\
&\le& c^2\mbox{Prob}(O_w=x|P_w=0...0).
\end{eqnarray*}

Let us assume that a language $L$ is in post-DQC1$_{n+1}$.
Then,
\begin{itemize}
\item[1.]
if $w\in L$ then $\mbox{Prob}(O_w'=1|P_w'=0...0)\ge\frac{1}{c^2}
(\frac{1}{2}+\delta)$
\item[2.]
if $w\notin L$ then $\mbox{Prob}(O_w'=1|P_w'=0...0)\le c^2
(\frac{1}{2}-\delta)$
\end{itemize}
If $c<\sqrt{2}$, $L$ is necessarily in post-BPP. 
(Note that it is assumed that $\delta$ can be made
arbitrarily close to 1/2. 
It is proved by noticing
the equivalence between post-DQC1$_{n+1}$
and post-BQP, and the coherent amplification of BQP acceptance probabilities).
This would imply that $\mbox{post-DQC1}_{n+1}\subseteq\mbox{post-BPP}$, and hence
$\Delta_3\text{P} \supseteq {\mbox{P}}^{\mbox{post-BPP}}
= {\mbox{P}}^{\mbox{PP}}
\supseteq \mbox{PH}$,
which shows the collapse of PH at the third level.
Hence we have shown that the classical simulability
of DQC1$_{n+1}$ leads to the collapse of PH
at the third level.

\begin{figure}[htbp]
\begin{center}
\[
\Qcircuit @C=1em @R=0.5em {
&\lstick{\ket{0}}&\qw & \multigate{5}{~W'~} & \qw & \qw &\meter\\
&\lstick{\frac{I}{2}}&\qw & \ghost{~W'~} & \qw & \targ & \meter \\
&\lstick{\frac{I}{2}}&\qw & \ghost{~W'~} & \gate{V_1} & \ctrl{-1}&\qw \\
&\push{\vdots} &&  &\push{\vdots} && \\
&\lstick{\frac{I}{2}}&\qw & \ghost{~W'~} & \gate{V_{n-1}} & \ctrl{-2}&\qw\\
&\lstick{\frac{I}{2}}&\qw & \ghost{~W'~} & \gate{V_n} & \qw &\meter
}
\]
\end{center}
\caption{
A circuit of the DQC1$_{3}$ model which implements postselected measurement based quantum computation.}
\label{DQC13}
\end{figure}
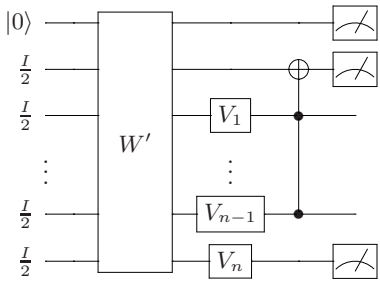

We now turn our attention to the case where measurements are available only on some constant number of output qubits. This more accurately reflects the situation in NMR experiments, as the DQC1 model was originally proposed to model. In order to show that post-BQP$\subseteq$post-DQC1$_3$ we consider the circuit shown in Fig \ref{DQC13}. Taking $W' = X \otimes |0\rangle\langle 0|\otimes |G\rangle\langle G| + I \otimes (I - |0\rangle\langle 0|\otimes |G\rangle\langle G|)$, the state after applying $W'$ and selecting on the first qubit being in the state $|1\rangle$ is $|0\rangle\otimes|G\rangle$, which is the same as in the previous proof with the addition of an ancilla qubit initialised to $|0\rangle$. As before, the local unitaries $\{V_i\}$ are used to align the local measurements which drive the computation with the computational basis. However, rather than directly measure each non-output qubit, the multiply-controlled Toffoli gate is used to compute the logical AND of these outcomes on the ancilla qubit. Thus, rather than postselecting on a particular string of outcomes, it suffices to postselect on the single ancilla bit in order to implement any postselected quantum circuit. Finally the output qubit must be measured in order to determine the result of the computation. Thus only three measurements, with post-selection on two of the output bits is necessary in order to implement post-selected quantum circuits, and hence post-BQP$\subseteq$post-DQC$1_3$. Following our previous argument, if DQC$1_3$ were classically simulable with multiplicative error less than $\sqrt{2}$, then post-BQP=post-DQC$1_3\subseteq$post-BPP, and hence 
PH would collapse at the third level.

In this paper, we have shown that classical efficient simulation
of DQC1$_k$ for any $k\geq 3$ is impossible unless PH collapses
at the third level. While we have derived this result using the measurement based model, we note that our results can also be recast in terms of the circuit model \cite{private}.
The ultimate goal is, of course, to show the impossibility
of a classical efficient simulation of DQC1. However, there is a link between these two problems. Shor and Jordan~\cite{SS} showed that
the probability of obtaining the all zero string result for a DQC$k_k$, where there are $k$ pure qubits which can be measured,
circuit is equal to the probability of obtaining the zero result
for a DQC1 circuit for trace estimation. Here we have shown that DQC1$_3$ is hard to classically simulated, and the Shor-Jordan result seems to indicate that the output bit of DQC1 somehow ``contains" the hard result of DQC1$_3$. While their result does not directly extend the post-selection argument to DQC1 circuits with a single measurement, it does show a tantalising link between the problems, and hence we are lead to conjecture that the result we present here can be extended to DQC1 circuits with only a single measurement. Such a result would be a major step toward resolving the computational power of DQC1 circuits.

We also note here that
the notion of approximate sampling used in the present paper (which is
also the one used in Refs.~\cite{BJS,AA})
is artificially strong. The natural notion of
classical simulation is to sample from a probability distribution of
1/poly total variation distance from the output distribution of the
device being simulated. 
It is an important open problem
to prove that
classical samplings with that natural error bound 
for a one clean qubit model (or other models~\cite{BJS,AA}) would
violate some plausible computational complexity assumptions.
Finally, we mention that due to the
postselected nature of the proof technique, we still do not know of a
non-promise problem solvable in probabilistic polynomial time by
DQC1$_k$ and plausibly not in classical probabilistic polynomial time.
Furthermore, it is not known whether the DQC1$_k$ is fault-tolerant, 
and it seems to be not.

We thank Michael Bremner, Animesh Datta, Oded Regev, Daniel Shepherd,
and two anonymous referees
for helpful comments on the manuscript. TM is supported by Tenure Track System by MEXT Japan.
KF is supported by JSPS Grant-in-Aid for
Research Activity Start-up 25887034.
JF is supported by the National Research Foundation and the
Ministry of Education, Singapore. This material is based on research funded
in part by the Singapore National Research Foundation under NRF Award NRF-NRFF2013-01.


\begin{thebibliography}{00}
\bibitem{KL}
E. Knill and R. Laflamme,
Phys. Rev. Lett. {\bf81}, 5672 (1998).
\bibitem{Poulin}
D. Poulin, R. Laflamme, G. J. Milburn, and J. P. Paz,
Phys. Rev. A {\bf68}, 022302 (2003).
\bibitem{Poulin2}
D. Poulin, R. Blume-Kohout, R. Laflamme,
and H. Ollivier,
Phys. Rev. Lett. {\bf92}, 177906 (2004).

\bibitem{earlier}
D. Shepherd, arXiv:0608132

\bibitem{SS}
P. W. Shor and S. P. Jordan,
Quant. Inf. Comput. {\bf8}, 681 (2008).
\bibitem{Datta2}
A. Datta, A. Shaji and C. M. Caves,
Phys. Rev. Lett. {\bf 100}, 050502 (2008).
\bibitem{Datta3}
A. Datta and G. Vidal,
Phys. Rev. A {\bf 75}, 042310 (2007).
\bibitem{Ambainis}
A. Ambainis, L. J. Schulman, and U. V. Vazirani,
Proc. of the 32nd Ann. ACM Sympo. on Theor. of Comput.
pp. 697 (2000).
\bibitem{Passante}
G. Passante, O. Moussa, C. A. Ryan, and R. Laflamme,
Phys. Rev. Lett. {\bf103}, 250501 (2009).
\bibitem{JW}
S. P. Jordan and P. Wocjan,
Quant. Inf. Comput. {\bf9}, 264 (2009).

\bibitem{3mani}
S. P. Jordan and G. Alagic, Proc. of the Sixth Conference
on Theory of Quantum Computation, Communication and Cryptography (TQC 2011);
arXiv:1105.5100

\bibitem{Datta}
A. Datta, S. T. Flammia, and C. M. Caves,
Phys. Rev. A {\bf72}, 042316 (2005).
\bibitem{TD}
B. Terhal and D. DiVincenzo,
Quant. Inf. Comput. {\bf4}, 134 (2004).
\bibitem{GC}
D. Gottesman and I. L. Chuang,
Nature {\bf402}, 390 (1999).
\bibitem{BJS}
M. J. Bremner, R. Jozsa, and D. J. Shepherd,
Proc. R. Soc. A {\bf467}, 2126 (2011).
\bibitem{AA}
S. Aaronson and A. Arkhipov,
Theory of Computing {\bf9}, 143 (2013).
\bibitem{FMIQP}
K. Fujii and T. Morimae,
arXiv:1311.2128
\bibitem{Aaronson}
S. Aaronson, 
Proc. R. Soc. A {\bf461}, 3437-3483 (2005).
\bibitem{KLM}
E. Knill, R. Laflamme, and G. J. Milburn,
Nature {\bf409}, 46 (2001).
\bibitem{Aaronson2}
S. Aaronson,
Proc. R. Soc. A {\bf467}, 2136 (2011).
\bibitem{MBQC}
R. Raussendorf
and H. J. Briegel, 
Phys. Rev. Lett. {\bf86}, 5188 (2001).
\bibitem{Toda}
S. Toda, 
SIAM J. Comput. {\bf20}, 865 (1991).
\bibitem{Barenco}
A. Barenco, C. H. Bennett, R. Cleve, D. P. DiVincenzo,
N. Margolus, P. Shor, T. Sleator, J. Smolin, and
H. Weinfurter,
Phys. Rev. A {\bf52}, 3457 (1995).
\bibitem{private}
M. Bremner and D. Shepherd, and separately Oded Regev,
private communication.

\bibitem{uniform}
As in the standard circuit computing, we consider a uniform family of circuits
in order to avoid any ``hard wiring" of hard computing.
In particular, as in Ref.~\cite{BJS}, we consider a mapping $w\to C_w$, where
$w$ is a bit string of length $n$, $C_w$ is a classical description of a circuit,
and the mapping is computable in classical poly($n$) time.
The description $C_w$ includes a specification of a sequance of gates and lines upon
which they act, a specification of the inputs for all lines, and a specification of
output registers and other registers such as the postselected ones.
\end{thebibliography}
\end{document}